\documentclass{svjour3}
\usepackage[utf8]{inputenc}
\usepackage[T1]{fontenc}
\usepackage{bbding}
\usepackage{graphicx}
\usepackage{amsmath,amssymb,amsfonts}
\usepackage{color}
\usepackage{algorithm}
\usepackage{algpseudocode}
\usepackage{algorithmicx}
\usepackage{verbatim}
\usepackage{tikz}
\usepackage{enumitem}
\usepackage{comment}
\usepackage{biblatex}
\usepackage{geometry}
\usepackage{hyperref}
\bibliography{refs}

\begin{document}

\title{A Framework for Initial Transient Detection and Statistical Assessment of Convergence in CFD Simulations}
\author{L. Scandurra$^1$ \and
        P. Alexias$^2$ \and
        E. de Villiers$^2$}

\institute{{\Envelope} L. Scandurra \at
			\email{l.scandurra@engys.com} \at \at
            {\Envelope} P. Alexias \at
            \email{p.alexias@engys.com} \at \at
            {\Envelope} E. de Villiers \at
            \email{e.devilliers@engys.com} \at \at
            $^{1}$ ENGYS Srl, Via del Follatoio 12, 34148 Trieste, Italy \\
            $^{2}$ ENGYS Ltd, London SW18 3SX, United Kingdom
}
\date{Received: date / Accepted: date}

\authorrunning{L. Scandurra, P. Alexias, E. de Villiers}
\titlerunning{SCA Method for CFD Simulations}

\maketitle

\begin{abstract}
Time series data often contain initial transient periods before reaching a stable state, posing challenges in analysis and interpretation. In this paper, we propose a novel approach to detect and estimate the end of the initial transient in time series data. Our method leverages the reversal mean standard error (RMSE) as a metric for assessing the stability of the data. Additionally, we employ fractional filtering techniques to enhance the detection accuracy by filtering out noise and capturing essential features of the underlying dynamics. 

Combining with autocorrelation-corrected confidence intervals we provide a robust framework to automate transient detection and convergence assessment.  The method ensures statistical rigor by accounting for autocorrelation effects, validated through simulations with varying time steps. Results demonstrate independence from numerical parameters (e.g., time step size, under-relaxation factors), offering a reliable tool for steady-state analysis. The framework is lightweight, generalizable, and mitigates inflated false positives in autocorrelated datasets.

\keywords{CFD \and Time series analysis \and Initial transient \and Statistical methods \and Confidence interval \and Convergence assessment}
\subclass{62M10, 62P30, 65C20, 65N12}
\end{abstract}

\section{Introduction}

Computational Fluid Dynamics (CFD) simulations, rely heavily on the analysis of time series data to ensure accuracy, stability, and meaningful statistical interpretation. Particularly, those models involving turbulent flows--whether using Reynolds-Averaged Navier-Stokes (RANS), Large Eddy Simulations (LES), or Direct numerical Simulation (DNS)--require careful examination of temporal behavior to capture transient phenomena and validate statistical convergence assessment \cite{bergmann2022statistical}. Detecting the initial transient phase is a critical challenge in CFD simulations, where the solution is dominated by numerical startup effects rather than physical behavior \cite{pasupathy2010initial}. If not handled properly, this phase can contaminate statistical averages, leading to erroneous conclusions.

The two main contributions of this work are transient identification to detect when the solution transitions from initial numerical artifacts to physically meaningful behavior, followed by data truncation to discard early-time data before statistical stationarity is achieved.

Beyond transient removal, another fundamental challenge is determining the optimal termination point of a simulation. Stopping too early risks incomplete convergence, while unnecessary prolongation wastes computational resources. Manual evaluation of these criteria is often subjective and inefficient, leaving the results vulnerable to statistical bias.

Automated convergence assessment addresses both challenges systematically: it not only identifies and discards transient distortions, but also determines the earliest point at which statistical stationary is achieved, ensuring simulations stop precisely when further computation yields diminishing returns.

With this motivation, we introduce in this article a data-driven approach that employs an original filtering algorithm to automatically detect where the initial transient phase ends and convergence begins.

To appreciate the necessity of such an automated method, it is important to understand current practices in the field, which often fall short of the required rigor \cite{mockett2010detection}. Typically, analysts resort to the following.
\begin{itemize}
    \item[$\bullet$] Visual inspection of trace plots, making a subjective judgment on where the solution appears "stable".
    \item[$\bullet$] Applying ad hoc rules, such as discarding a fixed percentage - such as 10\% or 20\% - of the initial samples.
    \item[$\bullet$] Simply running fixed-length chains and assuming convergence has been achieved without formal testing.
    \item[$\bullet$] While more advanced, case-specific heuristics or statistical comparisons exist, they are not universally applied.
\end{itemize}
However, these common methods come with significant limitations. Manual selection is inherently subjective and inefficient, risking the retention of biased data or the discarding of useful information \cite{wang2016marginal}. Furthermore, simplistic heuristics lack generality and are not reliable across diverse CFD cases.

The consequences are threefold. First, wasted computational resources from simulations that run longer than necessary. Second, subjective and biased results that can lead to erroneous conclusions. And third, an overall inefficient use of valuable computational time.

Therefore, an ideal solution must accomplish several key objectives. It needs to optimize run time by stopping simulations as soon as the results reach a defined confidence level. It must detect initial transient phases with precision to exclude non-representative data. Critically, it must account for autocorrelation, a fundamental characteristic of CFD time series where successive data points are not independent \cite{zikeba2010effective}. Ignoring this autocorrelation leads to a severe overestimation of the effective sample size and, consequently, inflated false positives in convergence diagnostics \cite{zikeba2011standard, melard1987confidence}.

Our proposed data-driven framework addresses these challenges by quantifying convergence through statistical moments like the mean and variance. The algorithm leverages the principle that newer data is the most representative of the true signal mean. A key innovation is a reverse-order standard error (RMSE) calculation, which processes the data starting from the latest sample and moving backwards in time to identify where the mean stabilizes. This is coupled with a filter designed to remove high-frequency noise and outliers, revealing the underlying trend.

The core of the method identifies a global minimum in this filtered RMSE. This global minimum signifies the point of smallest deviation from the established mean, marking the potential end of the transient. As the filter smooths the time series, local minima in the RMSE emerge as candidate points. A candidate is validated only if it satisfies a user-defined error threshold. The majority of validated candidates signify the most persistent global minimum position across all the filtering frequencies. 

The method provides a robust confidence interval assessment. The user specifies a confidence level $C$ and an error range $E$—for instance, requiring $95\%$ probability that the estimated mean is within $0.001$ of the true mean. To correctly handle autocorrelated data, this calculation uses an adjusted effective sample size $N_{\text{eff}}$ \cite{zikeba2010effective, zikeba2011standard}, which is significantly smaller than the total number of samples $N$, thereby providing statistically sound convergence criteria. Finally, to eliminate false positives from slow, low-frequency drifts, a final check is performed by fitting a trend-line to the data where convergence is confirmed only if the slope of this line is sufficiently small, ensuring true statistical stationarity.

Combining all these tools, the outcome is an instrument that delivers consistent convergence assessment, independent of numerical parameters such as time step size or Under-Relaxation Factor (URF), providing a reliable, automated, and objective standard for the CFD community.

Crucially, this tool is designed for accessibility, empowering CFD practitioners (even those without a deep statistical background) to obtain robust results with ease. The process is intentionally simplified: the user only needs to provide three intuitive inputs, the dataset to be analyzed, the desired confidence level, and an acceptable error range. With these parameters, the algorithm automatically executes the necessary statistical analysis and provides a clear, definitive result, making advanced convergence monitoring both straightforward and reliable.\\

The paper is organized as follows. A complete description of the proposed data-driven framework for automated convergence assessment is presented in Section 2. Section 3 details the original filtering algorithm designed to remove high-frequency noise and reveal the underlying trend in the time series. The core convergence assessment procedure, including the reverse-order standard error calculation and the criteria for identifying the end of the transient phase, is elaborated in Section 4. Section 5 explains the critical adjustment for autocorrelated data and the calculation of the effective sample size ($N_{\text{eff}}$) for robust confidence intervals. Finally, the performance of the method is demonstrated and validated against common practices in Section 6.

\section{Methodology}

\subsection{Initial Transient Detection}

Given a finite time series from a CFD simulation with N samples:
\[ X=\{ x_1,\ldots,x_N \} \]
we assume that the system reaches steady-state after some unknown time index $t_\text{cut}$, where $1\leqslant t_\text{cut}\leqslant N$. The goal is to determine $t_\text{cut}$, beyond which the data can be considered to come from a stationary distribution.

We define for every $i=1,\ldots,N$ the reverse cumulative mean $\bar{x}^\text{R}_i$, where the superscript 
$R$ indicates the Reversal component and computed from point $i$ to the end of the time series:
\[\bar{x}^\text{R}_i=\frac{1}{N-i+1}\displaystyle\sum^N_{j=i}x_j,\quad\text{for}\;i=1,2,\ldots,N\]

We also compute the RMSE, denoted for practical reasons as $\text{SEM}^\text{R}_i$, to quantify the uncertainty associated with each $\bar{x}^\text{R}_i$:
\begin{equation}\label{RMSE}
    \text{SEM}^\text{R}_i = \frac{s^\text{R}_i}{\sqrt{N - i + 1}}, \quad s^\text{R}_i = \sqrt{ \frac{1}{N - i} \sum_{j=i}^{N} \left( x_j - \bar{x}^\text{R}_i \right)^2 }
\end{equation}
This measure provides a localized estimate of variability in the trailing portion of the dataset. A stable reverse mean accompanied by a decreasing or low standard error is indicative of steady-state behavior.

\subsection{Detection Criterion: Global Minimum}

After computing the RMSE (\ref{RMSE}), we identify the point in time where the system transitions from transient behavior to steady-state by locating a global minimum in the RMSE curve.

This approach is based on the assumption that, as the system approaches steady-state, fluctuations in the mean decrease and variability stabilizes, resulting in a sustained reduction in the standard error of the mean.

\begin{figure}[h!]
    \centering
    \includegraphics[width=0.6\textwidth]{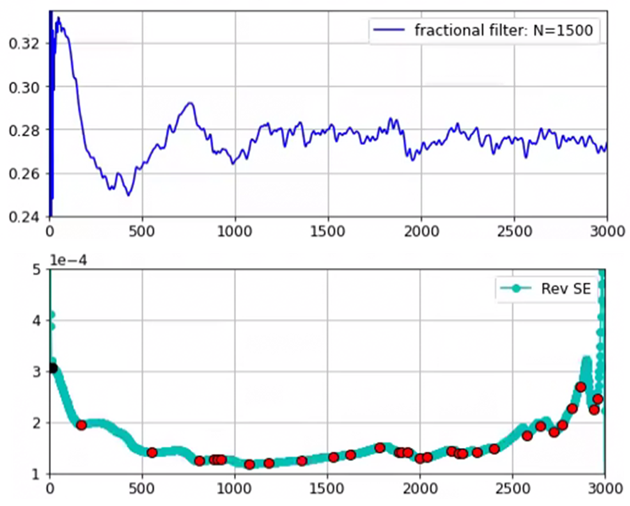} % Update with your actual image path
    \caption{Top: RANS time series dataset; Bottom: RMSE plot with potential local minima noted by red points.}
    \label{fig:FF_1}
\end{figure}
\newpage
However, due to residual noise or low-amplitude fluctuations, RMSE may exhibit multiple local minima that do not correspond to genuine changes in the underlying dynamics. To address this, a structured multiscale filtering technique based on a Fractional Filter (FF) is applied to the RMSE to suppress spurious variations while preserving the essential features of the signal.

The detection strategy follows a hierarchical approach. First, all local minima in the filtered RMSE curve are identified (Fig. \ref{fig:FF_1}). Each minimum is then evaluated based on the amplitude of the surrounding data, ensuring that only those occurring within regions of sufficient dynamic variation are retained.

\begin{figure}[h!]
    \centering
    \includegraphics[width=0.6\textwidth]{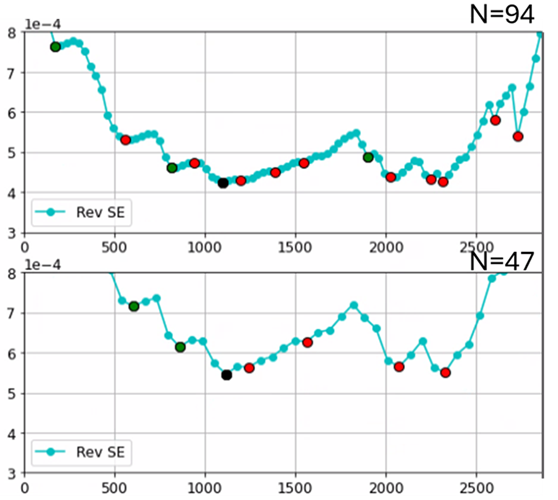} % Update with your actual image path
    \caption{RMSE graph with $N=94$ (top) and $N=47$ (bottom) number of samples. Red points represent all the local minima in the RMSE, Green points represent the local minima satisfies the error threshold and the black one the estimated global minimum.}
    \label{fig:FF_2}
\end{figure}

This condition is enforced using a threshold (user input) on the local range of the signal, which serves to eliminate minima that arise in flat or statistically insignificant portions of the series (Fig. \ref{fig:FF_2}).

Among the valid local minima, the one associated with the lowest RMSE value is selected as the most likely indicator of the end of the transient phase. The corresponding time index is denoted as $t_\text{cut}$ and defines the point beyond which the time series is considered to reflect statistically steady behavior.

\begin{figure}[h!]
    \centering
    \includegraphics[width=0.9\textwidth]{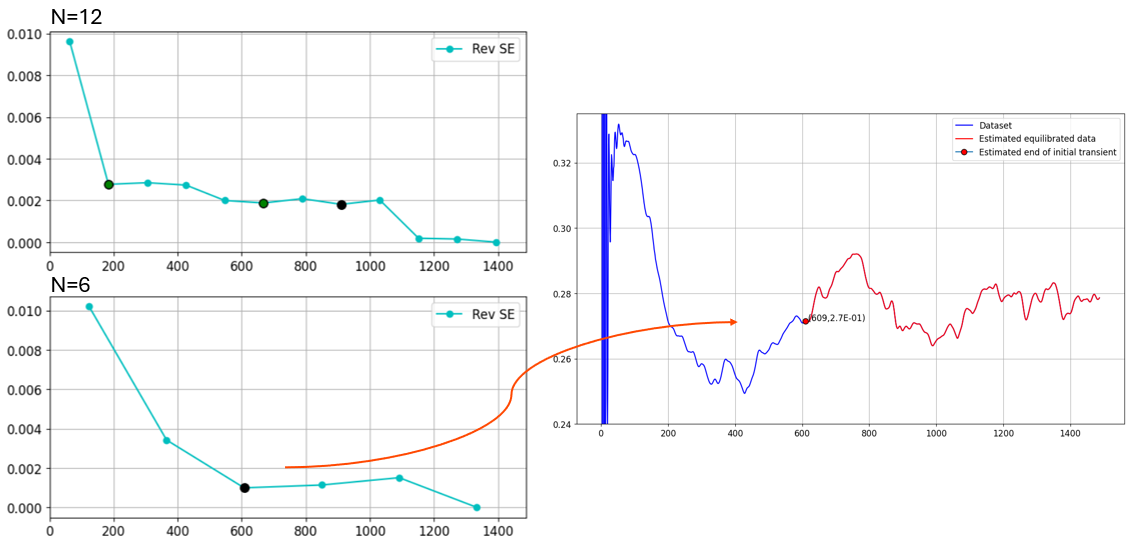} % Update with your actual image path
    \caption{Left: RMSE graph in both pictures with number of samples $N=12$ and $N=6$, respectively. The green points represent the local minima satisfies the error threshold and the black one the estimated global minimum. Right: the black point is mapped back to the original dataset.}
    \label{fig:FF_3}
\end{figure}

This value is then mapped back (Fig. \ref{fig:FF_3}) to the original dataset to locate the estimated transient endpoint. This criterion ensures that the estimation of $t_\text{cut}$ is both data-driven and robust. It avoids reliance on arbitrary window sizes or fixed thresholds, and adapts to the statistical structure of the specific time series under analysis.

% Fractional filtering (successive averaging with $N \rightarrow N/2$) smooths high-frequency noise while preserving trends. The global minimum in the filtered signal marks the transient endpoint.

\section{Fractional Filter Algorithm}
FF is a recursive multiscale smoothing technique designed to enhance signal structure and suppress high-frequency fluctuations without distorting the essential features of the data. Unlike traditional low-pass filters that operate at a single resolution, the FF reduces the signal resolution iteratively by combining adjacent elements to form a progressively coarser representation of the original sequence.
\newpage
A key advantage of this approach is its computational efficiency. The fractional filter is significantly less computationally expensive than a traditional low-pass filter, as it operates through simple averaging and downsampling rather than convolution with a long filter kernel or transformation to the frequency domain. This lower cost is the primary reason for its selection in our application, where performance is critical.

It is important to acknowledge that this efficiency comes with a trade-off: the fractional filter can suffer from more jitter in the smoothed output compared to a traditional low-pass filter. This is a direct result of the aggressive downsampling process. However, for the purpose of identifying broad trends and persistent features in the RMSE, this increased jitter is an acceptable compromise given the substantial reduction in computational cost.

\begin{figure}[h!]
    \centering
    \includegraphics[width=0.8\textwidth]{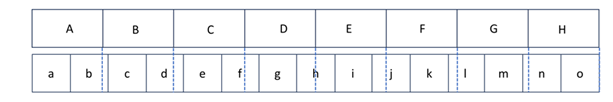} % Update with your actual image path
    \caption{Successive averaging $N=N/2$.}
    \label{fig:FF}
\end{figure}

At each stage of this downsampling process, the size of the signal is approximately halved (Fig. \ref{fig:FF}), and the resulting smoothed signal captures broader trends that may be obscured at the original resolution. This process is repeated until a minimal number of points remains, enabling analysis of the signal at increasingly abstract scales.

When applied to the RMSE, the filter gradually reveals persistent features in the variability of the signal, while reducing the influence of localized numerical noise. This multiscale representation allows for more reliable identification of local minima (Fig. \ref{fig:FF_2}) that reflect meaningful transitions in the system's statistical behavior.

The strength of the fractional filter lies in its ability to balance sensitivity, stability, and computational cost. It enhances the visibility of long-range trends while mitigating the impact of short-term fluctuations. In the context of transient detection, it provides a clear and stable framework for isolating the time at which the system reaches steady-state, even in the presence of simulation noise or small-scale perturbations.

By embedding this filtering strategy within the detection criterion, the method offers a robust and systematic tool for automated identification of steady-state conditions in time-dependent numerical data.

\begin{algorithm}[H]
\caption{Detection of Transient Cutoff Using Fractional Filter}\label{alg:cap}
\begin{algorithmic}[1]
\Require Preprocessed time series $X$ of length $N$, associated time vector $t$, user-defined threshold $\varepsilon$
\Ensure Estimated transient cutoff time $t_{\text{cut}}$
\State Initialize: $X_{\text{filtered}} \gets X$, $t_{\text{filtered}} \gets t$, $h \gets N$
\While{$h > 2$}
\State Smooth $X_{\text{filtered}}$ using the fractional filter to reduce length to $h/2$
\State Apply the same reduction to $t_{\text{filtered}}$
\State $h \gets \lfloor h / 2 \rfloor$
\State Compute RMSE for $X_{\text{filtered}}$
\State Identify all local minima in the RMSE sequence
\State Initialize empty set of valid minima $\mathcal{M}$
\For{each local minimum index $i$}
\State Compute local spread $\Delta_i = \max(X[i-1:i+2]) - \min(X[i-1:i+2])$
\If{$\Delta_i > \varepsilon$}
\State Add $i$ to $\mathcal{M}$
\EndIf
\EndFor
\If{$\mathcal{M}$ is not empty}
    \State Select index $i^{*}$ in $\mathcal{M}$ corresponding to minimum RMSE
\Else
    \State Select global minimum index $i^{*}$ in RMSE
\EndIf
\State $t_{\text{min}} \gets t_{\text{filtered}}[i^{*}]$
\EndWhile
\State \Return $t_{\text{cut}} = \lfloor t_{\text{min}} \rfloor$
\end{algorithmic}
\end{algorithm}
\noindent The pseudo-algorithm (Algorithm \ref{alg:cap}) summarizes this filtering and the detection process.

\section{Convergence Assessment}

Once the steady-state portion of the time series has been identified, it becomes essential to evaluate whether the results exhibit statistical convergence. This is accomplished by estimating a confidence interval for the mean of the steady-state data and comparing its width to a user-defined acceptable error threshold. For this final assessment, the diagnostic Reverse Mean Standard Error (RMSE) is replaced by the conventional Standard Error of the Mean (SEM) as the core metric of uncertainty. The SEM is calculated over the entire identified steady-state portion and is used to construct the confidence interval. If the uncertainty, quantified by the half-width of this interval, is sufficiently small, the time series can be considered statistically converged.

\subsection{Confidence Interval Estimation}

Let $X_\text{ss} = \{x_{t_\text{cut}}, x_{t_\text{cut}+1}, \ldots, x_N\}$ denote the portion of the time series after the estimated transient cutoff $t_\text{cut}$. We assume that this segment represents a stationary stochastic process, and we seek to quantify the uncertainty in its mean using a confidence interval.

Let $C$ be the confidence level specified by the user (e.g., $C = 0.95$), and define the significance level as $\alpha = 1 - C$. The two-tailed $t$-quantile is then evaluated at
\[
q = 1 - \frac{\alpha}{2}
\]
with degrees of freedom $\nu = n - 1$, where $n$ is the number of samples in $X_\text{ss}$. The $100 \times C\%$ confidence interval $\text{CI}$ for the sample mean $\bar{x}$ is computed as:
\[
\text{CI}=\bar{x} \pm t_{q,\ \nu} \cdot \text{SEM}
\]
where:
\begin{itemize}
    \item $\bar{x}$ is the sample mean of $X_\text{ss}$,
    \item $\text{SEM}$ is the standard error of the mean, given by
    \[
    \text{SEM} = \frac{s}{\sqrt{n}},
    \]
    with $s$ the sample standard deviation of $X_\text{ss}$,
    \item $t_{q,\ \nu}$ is the quantile of the Student's $t$-distribution corresponding to probability $q$ and $\nu$ degrees of freedom.
\end{itemize}

\subsection{Convergence Criterion}

To assess convergence, the width of the confidence interval is compared to a user-defined tolerance $\varepsilon$, which represents the maximum acceptable uncertainty in the estimated mean. The convergence condition is defined as:
\[
t_{q,\ \nu} \cdot \text{SEM} < \varepsilon
\]

If the condition is satisfied, the time series is deemed to have reached statistical convergence: the steady-state mean is known within the specified confidence level and acceptable error margin. Conversely, if the inequality is not met, the time series may still be affected by residual fluctuations or insufficient sampling, and additional simulation time may be required.

\subsection{Trend-Line Stability Check}

In addition to the confidence interval, a trend-line analysis is used to detect slow drifts in the time series mean. This guards against false positives in convergence due to long-term, low-frequency trends that may not be captured by statistical noise alone.

\begin{figure}[h!]
    \centering
    \includegraphics[width=0.4\textwidth]{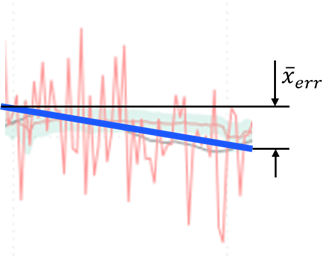} % Update with your actual image path
    \caption{Check whether the accumulated trend over all points ($n\times\text{slope}$) is less than the tolerance $\varepsilon$.}
    \label{fig:trend_base_conv}
\end{figure}

In Fig. \ref{fig:trend_base_conv} a linear regression is fitted to the steady-state data, and the slope $\partial\bar{x} / \partial n$ of the trend line is assessed. If the estimated mean continues to change systematically over time, this indicates a lack of convergence.

The following trend-based convergence condition is applied:
\[
n \cdot \left| \frac{\partial \bar{x}}{\partial n} \right| \leq \varepsilon
\]
This condition ensures that the expected change in the sample mean due to continued sampling is below the specified tolerance $\varepsilon$. If the trend is sufficiently flat, the system is considered stable in the mean.

\begin{comment}

Convergence is achieved when the confidence interval (CI) of the mean falls within a user-specified error range $E$:
\begin{equation}
    \text{CI} = \hat{\Theta} \pm t_{(1-\alpha/2,\nu)} \cdot \text{SE}, \quad \alpha = 1 - P
\end{equation}
where $P$ is the desired probability (e.g., 95\%) and $t$ is the Student's $t$-distribution critical value.
\end{comment}

\section{Autocorrelation Correction}

In convergence assessment, the estimation of confidence intervals is typically based on the assumption that the data points are independent and identically distributed \cite{bergmann2022statistical,melard1987confidence}. However, this assumption does not hold for many time series, especially those arising from physical simulations, iterative solvers, or under-relaxed numerical algorithms. In such cases, the data are often strongly autocorrelated — that is, each data point is statistically dependent on its predecessors.

This autocorrelation violates the independence assumption and leads to two major issues:

\begin{itemize}
    \item \textbf{Overconfident uncertainty estimates:} Confidence intervals underestimate the true uncertainty in the sample mean, increasing the risk of false positives.
    \item \textbf{Overestimation of the effective sample size:} The number of statistically independent observations is much smaller than the total number of samples, leading to premature or misleading convergence claims.
\end{itemize}

To address this, the number of effective independent samples \( N_{\text{eff}} \) must be estimated by accounting for the autocorrelation structure of the time series.

\subsection{Autocorrelation Function and Effective Sample Size}

Let \( X_\text{ss} = \{x_{t_\text{cut}}, x_{t_\text{cut}+1}, \ldots, x_N\} \) denote the steady-state portion of the time series after the transient has been removed. Assume this segment represents a weakly stationary stochastic process.

The degree of correlation between samples is quantified by the \textit{autocorrelation function} (ACF), defined for lag \( k \) as:

\begin{equation}
    \rho_k = \frac{\mathbb{E}\left[(x_t - \mu)(x_{t+k} - \mu)\right]}{\sigma^2}
\end{equation}
\noindent where, we recall, $\mu$ is the population mean and $\sigma$ is the population variance.

In practice, \( \rho_k \) is estimated using the sample mean \( \bar{x} \) and sample variance \( s^2 \), as:

\begin{equation}
    \rho_k \approx \frac{1}{(N - k)s^2} \sum_{t=1}^{N-k} (x_t - \bar{x})(x_{t+k} - \bar{x})
\end{equation}

The autocorrelation coefficient \( \rho_k \) quantifies how much information is shared between values \( k \) steps apart. When the time series is strongly autocorrelated, these coefficients decay slowly, implying fewer effective degrees of freedom.

The \textit{effective sample size} \( N_{\text{eff}} \) \cite{zikeba2010effective, zikeba2011standard} is then estimated as:

\begin{equation}
    N_{\text{eff}} = \frac{N}{1 + 2 \displaystyle\sum_{k=1}^{N-1} \frac{N-k}{N} \rho_k}
    \label{eq:neff}
\end{equation}

This correction adjusts for the information redundancy caused by correlated samples. When autocorrelation is strong, the denominator grows, and \( N_{\text{eff}} \ll N \), yielding a more conservative estimate of uncertainty.

\subsection{Adjusted Confidence Interval with Autocorrelation}

The standard error of the sample mean must also be corrected to reflect \( N_{\text{eff}} \) instead of \( N \). The adjusted standard error is:

\begin{equation}
    \text{SEM}_{\text{eff}} = \frac{s}{\sqrt{N_{\text{eff}}}}
\end{equation}

This leads to the corrected confidence interval for the sample mean:

\begin{equation}
    \text{CI} = \bar{x} \pm t_{q,\ \nu_{\text{eff}}} \cdot \text{SEM}_{\text{eff}}
\end{equation}

where:
\begin{itemize}
    \item \( t_{q,\ \nu_{\text{eff}}} \) is the Student’s \( t \)-quantile for confidence level \( q \) and degrees of freedom \( \nu_{\text{eff}} = N_{\text{eff}} - 1 \),
    \item \( s \) is the sample standard deviation,
    \item \( \bar{x} \) is the sample mean.
\end{itemize}

This correction ensures that the confidence interval properly reflects the statistical uncertainty in the presence of autocorrelation.

%Eugene wrote:"The last three bullet points of your conclusions don’t make sense."

% \subsection{Benefits of Autocorrelation Correction}

% Incorporating autocorrelation into convergence assessment improves the robustness and generality of statistical inference. Specifically, it ensures that:

% \begin{itemize}
%     \item Convergence is not overestimated due to misleading confidence intervals.
%     \item The results are consistent across simulations with different time step sizes, under-relaxation factors, or numerical solvers.
%     \item The confidence level specified by the user is preserved in practice, avoiding false convergence conclusions.
% \end{itemize}

% By correctly quantifying the effect of statistical dependence in the data, convergence criteria become more reliable and broadly applicable across computational setups.

% \begin{comment}

% Autocorrelation inflates false positives by reducing effective sample size. The adjustment is:
% \begin{equation}
%     N_{eff} = \frac{N}{1 + 2 \sum_{k=1}^{N-1} \frac{N-k}{N} \rho_k}
% \end{equation}
% where $\rho_k$ is the autocorrelation at lag $k$. For continuous signals, $\rho(\tau) = \exp(-\tau/T_0)$.
% \end{comment}

\section{Applications}

To validate the implementation of the proposed method, several test cases were analyzed, ranging from synthetic signals with known statistics to CFD-based signals obtained from RANS and DDES simulations. The following subsections summarize the outcomes and discuss the behavior observed in each case. %a set of synthetic signals with known statistical moments are used.

\subsection*{Random Gaussian Distribution}

For the first validation example, a synthetic signal is generated using a random Gaussian distribution with the following characteristics:
\begin{itemize}
    \item Mean Value: $0.3$
    \item Standard Deviation: $0.0066$
\end{itemize}

\begin{figure}[h!]
    \centering
    \includegraphics[width=0.7\textwidth]{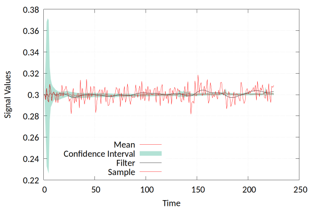} % Update with actual path
    \caption{Synthetic signal generated with Gaussian noise (Test 1).}
    \label{fig:test1_signal}
\end{figure}

Using a confidence interval of $99\%$ and a tolerance of $0.001$, the method converges after 225 samples and computes a mean value of $0.3004$ and a standard deviation of $0.0066$ (Table~\ref{tab:test1_stats}). As shown in Figure~\ref{fig:test1_signal}, the estimated mean quickly stabilizes around the target value as the sample size increases. The confidence interval decreases progressively, reflecting the reduction in uncertainty as more data is accumulated. This behavior confirms that the method correctly estimates both the mean and the associated uncertainty for purely random stationary signals.

%the evolution of the mean value and the confidence interval as the sample size increases.

\begin{table}[h!]
    \centering
    \caption{Test 1: Statistical evaluation of synthetic signal.}
    \begin{tabular}{|c|c|}
        \hline
        \textbf{Metric} & \textbf{Value} \\
        \hline
        Mean & 0.3004 \\
        \hline
        Standard Deviation & 0.0066 \\
        \hline
        Samples & 225 \\
        \hline
    \end{tabular}
    \label{tab:test1_stats}
\end{table}
\vspace{-1cm}
\subsection*{Random Gaussian Distribution With Transient Period}

In the second validation test, a transient part is added to the previous signal to simulate a system response or external influence, as commonly observed %is often seen 
in CFD applications. The transient %time
consists of a sinusoidal %signal
perturbation superimposed %with the underlined random 
on the Gaussian noise
%distribution and stops at 
and ends at approximately $200s$.

\begin{figure}[h!]
    \centering
    \includegraphics[width=0.7\textwidth]{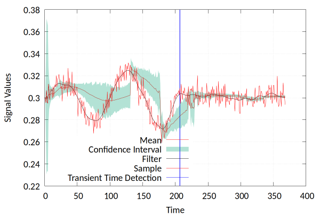} % Update with actual path
    \caption{Synthetic signal with transient addition (Test 2).}
    \label{fig:test2_signal}
\end{figure}

The proposed method successfully identifies this transient period, estimating its end at $207s$, very close to the actual termination time. Figure~\ref{fig:test2_signal} shows how the mean value and its confidence interval fluctuate significantly during the transient stage and then stabilize as the flow reaches a statistically steady regime. The predicted mean ($0.3012$) and standard deviation ($0.0064$) are again consistent with the expected stationary behavior (Table~\ref{tab:test2_stats}).

%As show in table \ref{tab:test2_stats}, this time, the algorithm predicts a mean value of $0.3012$ and a standard deviation of $0.064$. It also estimates that the transient time stops at $207s$ which is very close to the actual value of $200s$. Figure \ref{fig:test2_signal} shows the evolution of the mean value and the confidence interval together with the transient time detection estimation. 

\begin{table}[h!]
    \centering
    \caption{Test 2: Statistical evaluation with transient part.}
    \begin{tabular}{|c|c|}
        \hline
        \textbf{Metric} & \textbf{Value} \\
        \hline
        Mean & 0.3012 \\
        \hline
        Standard Deviation & 0.0064 \\
        \hline
        Samples & 162 \\
        \hline
        Transient Time (s) & 207 \\
        \hline
    \end{tabular}
    \label{tab:test2_stats}
\end{table}

\subsection{Example: RANS Evaluation – External Aero Case}

\label{sec:rans_external_aero}

%This result presents the performance of the RANS solver in an external aerodynamics scenario. The signal used for monitoring was normalized, and the analysis was carried out with a target interval of $0.001$ under a $95\%$ confidence level.

This case evaluates the performance of the RANS solver in an external aerodynamics scenario. The monitored signal—normalized for analysis—represents a typical convergence trace where transient startup effects gradually vanish.

\begin{figure}[h!]
    \centering
    \includegraphics[width=0.75\textwidth]{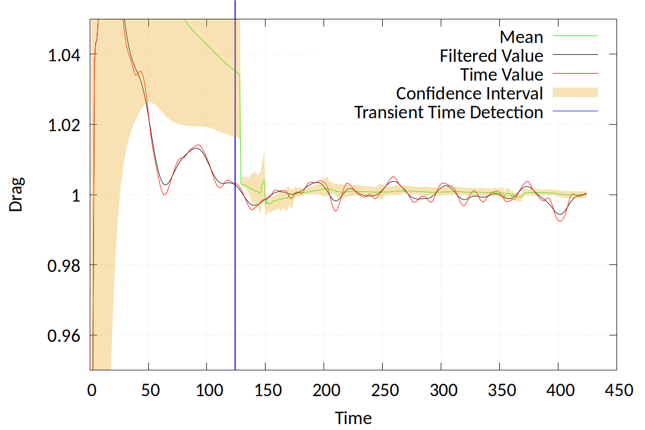} % Replace with actual path
    \caption{Normalized signal convergence behavior in RANS external aero simulation.}
    \label{fig:rans_signal}
\end{figure}

\begin{table}[h!]
    \centering
    \caption{RANS simulation performance metrics.}
    \begin{tabular}{|c|c|}
        \hline
        \textbf{Metric} & \textbf{Value} \\
        \hline
        Iterations to Converge & 424 \\
        \hline
        Transient Iterations & 147 \\
        \hline
        Full Signal Iterations & 590 \\
        \hline
        Error & 0.015\% \\
        \hline
        CPU Saving & 28\% \\
        \hline
    \end{tabular}
    \label{tab:rans_metrics}
\end{table}

Figure~\ref{fig:rans_signal} illustrates the signal’s evolution: an initial oscillatory behavior characterizes the early transient phase, followed by a clear stabilization trend as the mean value converges toward a constant level. The confidence bounds shrink accordingly, indicating increased statistical reliability.

The statistical convergence algorithm detects convergence after $424$ iterations, while the total recorded signal extends to $590$ iterations (Table~\ref{tab:rans_metrics}). The early termination of the simulation would therefore provide a CPU time saving of $28\%$ without compromising accuracy (error of only $0.015\%$).

This result highlights the potential of the method to reduce computational cost in steady-state RANS simulations, particularly for cases where traditional residual-based convergence criteria are too conservative.

\subsection{DDES Evaluation – GMT (Truck)}
\label{sec:ddes_gmt_truck}

This example applies the method to a Delayed Detached Eddy Simulation (DDES) for the GMT (truck) case.

%The analysis is performed using a drag signal with a statistical approach targeting an error of $0.001$ with a $95\%$ confidence level.

Unlike the RANS case, this unsteady flow exhibits significant temporal fluctuations due to large-scale vortex shedding and turbulent wake dynamics.

\begin{figure}[h!]
    \centering
    \includegraphics[width=0.75\textwidth]{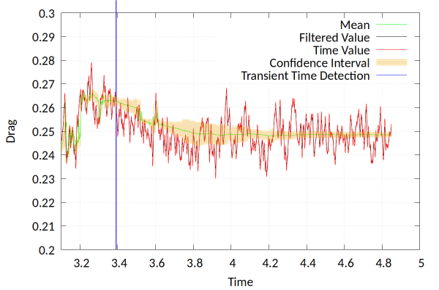} 
    \caption{Full normalized signal from GMT (Truck) DDES simulation.}
    \label{fig:ddes_signal_full}
\end{figure}

In Figure~\ref{fig:ddes_signal_full}, the full normalized drag coefficient signal is shown. The early portion of the trace displays a pronounced transient response as the flow develops, with high-amplitude oscillations gradually settling into a statistically stationary pattern.

\begin{figure}[h!]
    \centering
    \includegraphics[width=0.75\textwidth]{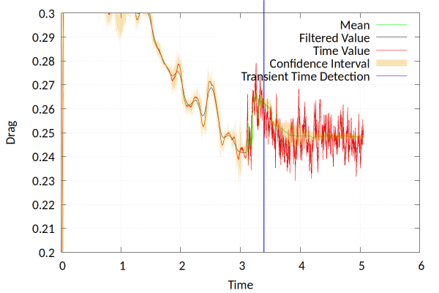} 
    \caption{Zoomed-in view highlighting the end of the initial transient phase.}
    \label{fig:ddes_signal_zoom}
\end{figure}

The zoomed-in view in Figure~\ref{fig:ddes_signal_zoom} highlights this transition more clearly: after approximately $3.4s$, the oscillations maintain a stable mean and amplitude, suggesting that the system has reached its quasi-steady turbulent regime.

\begin{table}[h!]
    \centering
    \caption{DDES simulation performance summary for GMT (Truck) case.}
    \begin{tabular}{|c|c|}
        \hline
        \textbf{Metric} & \textbf{Value} \\
        \hline
        Time for Convergence & 4.84883 sec \\
        \hline
        Transient Time & 3.39242 sec \\
        \hline
        Best Practice & 5.1 sec \\
        \hline
        SCA Mean Value & 0.248439 \\
        \hline
        Best Practice Mean & 0.248048 \\
        \hline
        Error & 0.15\% \\
        \hline
        CPU Saving & 9.97\% \\
        \hline
    \end{tabular}
    \label{tab:ddes_gmt_stats}
\end{table}

Quantitatively, the method detects a transient period lasting $3.392s$, in close agreement with the manually determined best-practice value of $5.1s$ (Table~\ref{tab:ddes_gmt_stats}). The predicted mean drag coefficient ($0.248439$) differs by only $0.15\%$ from the reference ($0.248048$), confirming the robustness of the approach. The estimated CPU saving of about $10\%$ further demonstrates its efficiency in unsteady, high-fidelity simulations.

Overall, the DDES results underline the method’s capability to handle complex, highly correlated flow signals while maintaining accurate statistical estimations. The detection of the transient endpoint and reliable convergence of the mean signal provide strong evidence that the approach can be generalized to other unsteady aerodynamic applications.

\subsection*{Summary Interpretation}

Across all test cases—from synthetic signals to realistic CFD data—the proposed SCA method consistently identifies the point at which the monitored signal becomes statistically stationary.

\begin{itemize}
    \item[$\bullet$] In the RANS case, this results in significant computational savings with negligible error.
    \item[$\bullet$] In the DDES case, despite the higher complexity and turbulence-induced variability, the algorithm maintains accuracy and still achieves measurable time savings.
\end{itemize}
These findings confirm the potential of SCA as an adaptive, data-driven convergence criterion applicable to both steady and unsteady CFD simulations.

\printbibliography

@article{bergmann2022statistical,
  author    = {Bergmann, Michael and Morsbach, Christian and Ashcroft, Graham and K{\"u}geler, Edmund},
  title     = {Statistical error estimation methods for engineering-relevant quantities from scale-resolving simulations},
  journal   = {Journal of Turbomachinery},
  volume    = {144},
  number    = {3},
  pages     = {031005},
  year      = {2022},
  publisher = {American Society of Mechanical Engineers}
}

@inproceedings{mockett2010detection,
  author       = {Mockett, Charles and Knacke, Thilo and Thiele, Frank},
  title        = {Detection of initial transient and estimation of statistical error in time-resolved turbulent flow data},
  booktitle    = {Proceedings of the 8th International Symposium on Engineering Turbulence Modelling and Measurements},
  pages        = {9--11},
  year         = {2010},
  organization = {European Research Collaboration on Flow Turbulence and Combustion Marseille~…}
}

@article{zikeba2010effective,
  author    = {Zi{\k{e}}ba, Andrzej},
  title     = {Effective number of observations and unbiased estimators of variance for autocorrelated data-an overview},
  journal   = {Metrology and Measurement Systems},
  number    = {1},
  year      = {2010},
  publisher = {Polska Akademia Nauk}
}

@article{zikeba2011standard,
  author    = {Zi{\k{e}}ba, Andrzej and Ramza, Piotr},
  title     = {Standard deviation of the mean of autocorrelated observations estimated with the use of the autocorrelation function estimated from the data},
  journal   = {Metrology and Measurement Systems},
  volume    = {18},
  number    = {4},
  pages     = {529--542},
  year      = {2011}
}

@inproceedings{pasupathy2010initial,
  author       = {Pasupathy, Raghu and Schmeiser, Bruce},
  title        = {The initial transient in steady-state point estimation: Contexts, a bibliography, the MSE criterion, and the MSER statistic},
  booktitle    = {Proceedings of the 2010 Winter Simulation Conference},
  pages        = {184--197},
  year         = {2010},
  organization = {IEEE}
}

@article{wang2016marginal,
  author    = {Wang, Rob J and Glynn, Peter W},
  title     = {On the marginal standard error rule and the testing of initial transient deletion methods},
  journal   = {ACM Transactions on Modeling and Computer Simulation (TOMACS)},
  volume    = {27},
  number    = {1},
  pages     = {1--30},
  year      = {2016},
  publisher = {ACM New York, NY, USA}
}

@article{melard1987confidence,
  title     = {On confidence intervals and tests for autocorrelations},
  author    = {Melard, Guy and Roy, Roch},
  journal   = {Computational Statistics \& Data Analysis},
  volume    = {5},
  number    = {1},
  pages     = {31--44},
  year      = {1987},
  publisher = {Elsevier}
}

\end{document}